\documentclass[aps,prb,showpacs,reprint]{revtex4-1}
\usepackage{graphicx}

\begin{document}


\title{Thin films versus 2D sheets in layered structures: graphene and 2D metallic sheets} 



\author{Bo E. Sernelius}
\email[]{bos@ifm.liu.se}
\affiliation{Division of Theory and Modeling, Department of Physics, Chemistry, and Biology, Link{\" o}ping University, SE-581 83 Link{\" o}ping, Sweden.}


\date{\today}

\begin{abstract}	
We study an interface between two media separated by a strictly 2D sheet. We show how the amplitude reflection coefficient  can be modeled by that for an interface where the 2D sheet has been replaced by a film of small but finite thickness. We give the relationship between the 3D dielectric function of the thin film and the 2D dielectric function of the sheet. We choose graphene and a 2D metallic sheet as illustrative examples. This approach turns out to be very useful when treating graphene or graphene like sheets in non-planar structures.
\end{abstract}


\maketitle 

\section{Introduction}

There are no strictly 2D (two-dimensional) systems in the real world. However, if the carriers are strongly confined in one direction they have quantized energy levels for one spatial dimension but are free to move in two spacial dimensions\cite{Ando}. Thus the wave vector is a good quantum number for two dimensions but not for the third. Examples are narrow quantum wells and graphene. In many situations these systems can be approximated by the idealized systems of strictly 2D sheets. Even thicker films where the spatial quantization is negligible behave as strictly 2D from distances much larger than the film thickness. The dielectric functions of a 2D electron gas\cite{Stern} and for graphene\cite{Guinea,Wun,Sarma,Sera,Ser1} have been derived in the literature. The dielectric function for graphene is valid for an idealized 2D sheet. If one were to treat graphene as a film of finite thickness one could not use the 2D dielectric function as it is. It has to be modified into a 3D version. The same thing applies to a strictly 2D metal sheet if it were to be treated as a film of finite thickness. If a metal film of finite thickness with a bulk dielectric function were to be treated as a strictly 2D sheet the dielectric function had to be modified into a 2D version. How these modifications can be done is what this paper is all about. Examples where these results are useful are for optical properties in layered structures and for dispersion interactions (van der Waals and Casimir) in layered structures.

In Sec. \ref{empty} we give the amplitude reflection coefficients for an interface between two media, in Sec. \ref{sheet} we show how these are modified when a 2D sheet is inserted at the interface, in Sec. \ref{film} we show how these are modified when instead a film of finite thickness is inserted, and in Sec. \ref{simul} we show how a 2D sheet can be modeled by a film of finite thickness.
In Secs. \ref{graphene} and \ref{metal} we illustrate this modeling for graphene, and a metal sheet, respectively. Before we end with a brief summary and conclusion section, Sec. \ref{sum}, we discuss in Sec. \ref{nonplanar} how the results for 2D sheets and thin films can be used in non-planar structures.
\section{Amplitude reflection coefficient at an interface\label{empty}}

For the present task we need a geometry consisting of two regions and one interface, $i|j$. For planar structures there are two types of mode, transverse magnetic (TM) or $p$-polarized and transverse electric (TE) or $s$-polarized. They have different amplitude reflection coefficients. At an interface between medium $i$ and $j$ the TM and TE amplitude reflection coefficients for waves impinging  from the $i$ side are
\begin{equation}
r_{ij}^{TM} = \frac{{{{\tilde \varepsilon }_j}{\gamma _i} - {{\tilde \varepsilon }_i}{\gamma _j}}}{{{{\tilde \varepsilon }_j}{\gamma _i} + {{\tilde \varepsilon }_i}{\gamma _j}}},
\label{equ1}
\end{equation}
and
\begin{equation}
r_{ij}^{TE} = \frac{{\left( {{\gamma _i} - {\gamma _j}} \right)}}{{\left( {{\gamma _i} + {\gamma _j}} \right)}},
\label{equ2}
\end{equation}
respectively.  Note that ${r_{ji}} =  - {r_{ij}}$ holds for both mode types. If retardation is neglected there is only one mode type and the amplitude reflection coefficient is
\begin{equation}
r_{ij} = \frac{{{{\tilde \varepsilon }_j} - {{\tilde \varepsilon }_i}}}{{{{\tilde \varepsilon }_j}+ {{\tilde \varepsilon }_i}}}.
\label{equ3}
\end{equation}
In the above equations ${\gamma _i} = \sqrt {1 - {{\tilde \varepsilon }_i}\left( \omega  \right){{\left( {\omega /ck} \right)}^2}} $, ${{{\tilde \varepsilon }_i}\left( \omega  \right)}$ is the dielectric function of medium $i$, $c$ is the speed of light in vacuum, and $k$ is the length of a wave vector in the plane of the interface. We have suppressed the arguments of the functions in Eqs. (\ref{equ1}-\ref{equ3}). The amplitude reflection coefficients and the $\gamma$ -functions are functions of $k$ and $\omega$. The dielectric functions are functions of $\omega$, only, i.e., spatial dispersion is neglected. Inclusion of spatial dispersion in the bulk dielectric functions is possible\cite{Ser3,EsqSvet} but would lead to much higher complexity and negligible effects in most cases.
\subsection{Interface with a strictly 2D sheet\label{sheet}}
There are different formulations of electromagnetism in the literature. The difference lies in how the conduction carriers are treated. In one formulation these carriers are lumped together with the external charges to form the group of free charges. Then only the bound charges contribute to the screening. We want to be able to treat geometries with metallic regions. Then this formulation is not suitable. In the formulation that we use the conduction carriers are treated on the same footing as the bound charges. Thus, both bound and conduction charges contribute to the dielectric function. In the two formalisms the {\bf E} and {\bf B} fields, the true fields, of course are the same. However, the auxiliary fields  the {\bf D} and {\bf H} fields are different. To indicate that we use this alternative formulation we put a tilde above the {\bf D} and {\bf H} fields and also above the dielectric functions. See Ref.[\onlinecite{Sera}] for a fuller discussion on this topic. 

Within this formalism the standard boundary conditions, used to derive the reflection coefficients, are that in absence of external charge and current densities at an interface the tangential components of the $\bf E$ and $\bf \tilde H$ fields and the normal components of the $\bf \tilde D$ and $\bf B$ fields are all continuous across the interface. The sources to the fields in Maxwell's equations are the external charge and current densities. In the boundary conditions any discontinuities in the normal component of the $\bf \tilde D$ fields and tangential component of the $\bf \tilde H$ fields are caused by external surface charge densities and external surface current densities, respectively. 

The amplitude reflection coefficient gets modified if there is a 2D layer at the interface. We treat\cite{Sera} the 2D layer at the interface as external to our system.  The modified amplitude reflection coefficient for a TM mode is \cite{Sera}
\begin{equation}
r_{ij}^{TM} = \frac{{{{\tilde \varepsilon }_j}{\gamma _i} - {{\tilde \varepsilon }_i}{\gamma _j} + 2{\gamma _i}{\gamma _j}{\alpha ^\parallel }}}{{{{\tilde \varepsilon }_j}{\gamma _i} + {{\tilde \varepsilon }_i}{\gamma _j} + 2{\gamma _i}{\gamma _j}{\alpha ^\parallel }}},
\label{equ4}
\end{equation}
where the polarizability, $\alpha^\parallel$, of the 2D sheet is obtained from the dynamical conductivity, $\sigma ^\parallel$,

\begin{equation}
{\alpha ^\parallel }\left( {k,\omega } \right) = \frac{{2\pi i{\sigma ^\parallel }\left( {k,\omega } \right)k}}{\omega },
\label{equ5}
\end{equation}
and the dielectric function is
\begin{equation}
{\varepsilon ^\parallel }\left( {k,\omega } \right) = 1 + {\alpha ^\parallel }\left( {k,\omega } \right).
\label{equ6}
\end{equation}
For TM modes the tangential component of the electric field, which will induce the external current, is parallel to {\bf k}, so the longitudinal 2D dielectric function of the sheet enters. The bound charges in the 2D sheet also contribute to the dynamical conductivity and the polarizability.

The modified amplitude reflection coefficient for a TE mode is \cite{Sera}
\begin{equation}
r_{ij}^{TE} = \frac{{{\gamma _i} - {\gamma _j} + 2{{\left( {\omega /ck} \right)}^2}{\alpha ^ \bot }}}{{{\gamma _i} + {\gamma _j} - 2{{\left( {\omega /ck} \right)}^2}{\alpha ^ \bot }}},
\label{equ7}
\end{equation}
where the polarizability, $\alpha ^ \bot $, of the 2D sheet is obtained from the dynamical conductivity, $\sigma ^ \bot $,
\begin{equation}
{\alpha ^ \bot }\left( {k,\omega } \right) = \frac{{2\pi i{\sigma ^ \bot }\left( {k,\omega } \right)k}}{\omega },
\label{equ8}
\end{equation}
and the dielectric function is
\begin{equation}
{\varepsilon ^ \bot }\left( {k,\omega } \right) = 1 + {\alpha ^ \bot }\left( {k,\omega } \right).
\label{equ9}
\end{equation}
For a TE wave the electric field is perpendicular to {\bf k}, so the transverse 2D dielectric function of the sheet enters. 
The bound charges in the 2D sheet also contribute to the dynamical conductivity and the polarizability.

If retardation is neglected there is only one mode type and the amplitude reflection coefficient is
\begin{equation}
r_{ij} = \frac{{{{\tilde \varepsilon }_j} - {{\tilde \varepsilon }_i} + 2{\alpha ^\parallel }}}{{{{\tilde \varepsilon }_j} + {{\tilde \varepsilon }_i} + 2{\alpha ^\parallel }}}.
\label{equ10}
\end{equation}
Now we have in Eqs. (\ref{equ4}), (\ref{equ7}), and (\ref{equ10}) the amplitude reflection coefficients for an interface between two media with a 2D sheet sandwiched in between. To be noted is that spatial dispersion of the 2D sheet can be included without any complications. This spatial dispersion has furthermore important effects as we will show later. In next section we will show the corresponding results when instead of a 2D sheet we have a thin film sandwiched between the two media.
\subsection{Interface with a film of finite thickness\label{film}}
For the present task we need a geometry consisting of three regions and two interfaces, $i|j|k$.
For this composite interface the amplitude reflection coefficient  for a wave impinging from the $i$ side is\,\cite{Maha, Ser0} 
\begin{equation}
{r_{ijk}} = \frac{{{r_{ij}} + {e^{ - 2{\gamma _j}k{d_j}}}{r_{jk}}}}{{1 + {e^{ - 2{\gamma _j}k{d_j}}}{r_{ij}}{r_{jk}}}},
\label{equ11}
\end{equation}
where $d_j$ is the thickness of the film $j$. This expression is valid for TM- and TE-modes when retardation is included and also for the modes when retardation is neglected. The appropriate amplitude reflection coefficient from Eqs. (\ref{equ1}-\ref{equ3}) should be used in the expression on the right hand side. In next section we show how the effect of a strictly 2D sheet can be modeled by a film of finite thickness with the proper choice of 3D dielectric function.

\subsection{Simulating a strictly 2D film with a film of finite thickness\label{simul}}
Let us study a 2D-sheet placed in a time varying electric field, $\bf E$, in the plane of the film. There will be a surface current density, ${\bf{K}} = {\sigma ^{2D}}{\bf{E}}$. This current density has contributions also from bound charges. Since we want to treat this 2D sheet as a thin film of finite thickness, $\delta $, we let this current be spread evenly through the thickness of the film. The volume current density, $\bf j$, is then
 ${\bf{j}} = {\bf{K}}/\delta $, and  since ${\bf{j}} = {\sigma ^{3D}}{\bf{E}}$ it follows that
 \begin{equation}
{\sigma ^{3D}} = {\sigma ^{2D}}/\delta.
\label{equ12}
\end{equation}
Now, since 
\begin{equation}
{\alpha ^{2D}} = 2\pi i{\sigma ^{2D}}k/\omega ,
\label{equ13}
\end{equation}
and
 \begin{equation}
{\alpha ^{3D}} = 4\pi i{\sigma ^{3D}}/\omega,
\label{equ14}
\end{equation}
we find that
\begin{equation}
{\alpha ^{3D}} = 2{\alpha ^{2D}}/k\delta .
\label{equ15}
\end{equation}
Thus in problems with 2D sheets one can treat the sheets as thin 3D films where the 3D polarizability above is used. To check if this is a reasonable approach we insert the expressions in Eqs. (\ref{equ1}-\ref{equ3}) into Eq. (\ref{equ11}) where now $d_j$ is $\delta$ and ${\tilde \varepsilon }_j=1+2{\alpha ^{2D}}/k\delta$. If we now let the film thickness, $\delta$, go towards zero we reproduce the results in Eqs. (\ref{equ4}), (\ref{equ7}), and (\ref{equ10}). Thus in the limit the model is exact.


\section{Graphene as an illustrative example\label{graphene}}
We will now calculate the Casimir interaction energy between two free standing undoped graphene sheets in vacuum. To make it as simple as possible we neglect retardation and perform the calculations for zero temperature. Retardation effects are actually very small in graphene \cite{Santos,Sera}. 
In a general point, $z$, in the complex frequency plane, away from the real axis the polarizability is \cite{Guinea}
\begin{figure}
\includegraphics[width=7.0cm]{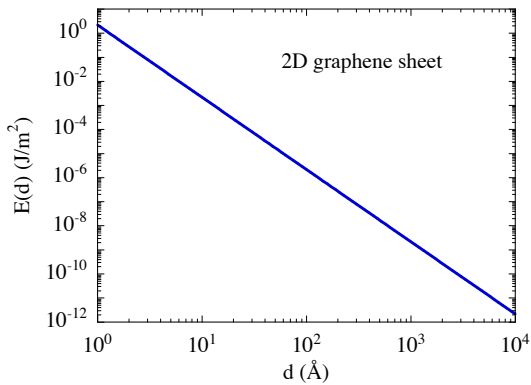}
\caption{(Color online) The attractive nonretarded\cite{Ser1} interaction energy per unit area between two graphene sheets as function of separation.}
\label{figu1}
\end{figure}

\begin{equation}
{\alpha ^{2D}}\left( {k,z} \right) = \frac{{2\pi {e^2}g}}{{16\hbar }}\frac{k}{{\sqrt {{v^2}{k^2} - {z^2}} }},
\label{equ16}
\end{equation}
where $v$ is the carrier velocity which is a constant in graphene ($E =  \pm \hbar vk$), and $g$ represents the degeneracy parameter with the value of 4 (a factor of 2 for spin and a factor of 2 for the cone degeneracy). In the numerical calculations we use the value\,\cite{Wun}  $8.73723 \times {10^5}$ m/s for $v$. If we now treat the graphene sheet as a thin film of thickness $\delta$ the polarizability of the film material should be chosen as
\begin{equation}
{\alpha ^{3D}}\left( {k,z} \right) = \frac{{\pi {e^2}}}{{\hbar \delta }}\frac{1}{{\sqrt {{v^2}{k^2} - {z^2}} }},
\label{equ17}
\end{equation}
and on the imaginary frequency axis it is
\begin{equation}
{\alpha ^{3D}}\left( {k,i\omega } \right) = \frac{{\pi {e^2}}}{{\hbar \delta }}\frac{1}{{\sqrt {{v^2}{k^2} + {\omega ^2}} }}.
\label{equ18}
\end{equation}
The van der Waals (vdW) and Casimir interactions can be derived in many different ways. 
One way is to derive the interaction in terms of the electromagnetic normal modes\cite{Ser0} of the system. For planar structures the interaction energy per unit area can be written as \cite{Ser0}
\begin{equation}
E = \hbar \int {\frac{{{d^2}k}}{{{{\left( {2\pi } \right)}^2}}}} \int\limits_0^\infty  {\frac{{d\omega }}{{2\pi }}} \ln \left[ {{f_k}\left( {i\omega } \right)} \right],
\label{equ19}
\end{equation}
where ${f_k}\left( {{\omega _k}} \right) = 0$ is the condition for electromagnetic normal modes. For the present geometry the mode condition function is
\begin{equation}
{f_k} = 1 - {e^{ - 2kd}}{\left( {{r_{121}}} \right)^2},
\label{equ20}
\end{equation}
where the index $1$ stands for vacuum and the index $2$ for the film material.
Using Eq. (\ref{equ11}) with the proper functions for our problem inserted we get
\begin{equation}
{r_{121}} = \frac{{{r_{12}} + {e^{ - 2k\delta }}{r_{21}}}}{{1 + {e^{ - 2k\delta }}{r_{12}}{r_{21}}}} = \frac{{{r_{12}}\left( {1 - {e^{ - 2k\delta }}} \right)}}{{1 - {e^{ - 2k\delta }}r_{12}^2}},
\label{equ21}
\end{equation}
where
\begin{equation}
{r_{12}} = \frac{{{\alpha ^{3D}}\left( {k,i\omega } \right)}}{{{\alpha ^{3D}}\left( {k,i\omega } \right) + 2}}.
\label{equ22}
\end{equation}
For strictly 2D sheets the corresponding mode condition function is
\begin{equation}
{f_k} = 1 - {e^{ - 2kd}}{\left[ {\frac{{{\alpha ^{2D}}\left( {k,i\omega } \right)}}{{1 + {\alpha ^{2D}}\left( {k,i\omega } \right)}}} \right]^2}.
\label{equ23}
\end{equation}
The result of Eq. (\ref{equ19}) with the mode condition function from Eq. (\ref{equ23}) is shown in Fig. \ref{figu1}. With the particular screening in graphene it turns out that $\alpha^{2D} \left( {k/\lambda ,i\omega /\lambda } \right) = \alpha \left( {k,i\omega } \right)$ and the separation dependence of the nonretarded interaction becomes very simple. A change in dummy variables removes the only $d$ in the integrand and produces the factor of ${d^{ - 3}}$ in front of the integral. The result is a straight line in Fig.\,\ref{figu1}. 

\begin{figure}
\includegraphics[width=7.0cm]{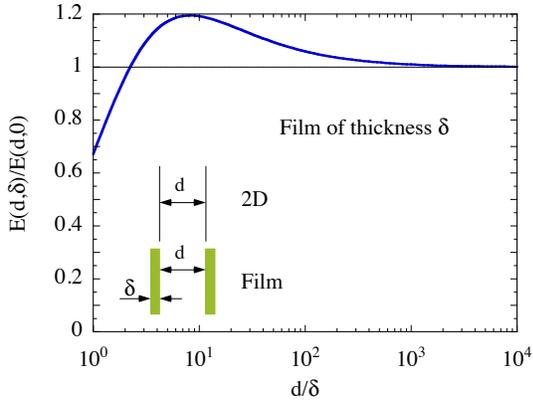}
\caption{(Color online) The ratio between the film result and the 2D sheet result for the interaction energy between two graphene sheets as function of the ratio between the separation and the film thickness.}
\label{figu2}
\end{figure}

When we treat the graphene sheets as thin films of thickness $\delta$ the momentum, $k$, enters in more places in the integrand with the effect that the result has a more complicated dependence on $d$. However in those additional places $k$ always enters in the combination $k\delta$. This means that there will be a universal correction factor, from treating the sheet as a film, that depends on $d/\delta$. This correction factor is shown in Fig. \ref{figu2}. One would expect the interaction to decrease for small separations since all matter in the two films is further apart than $d$, see the inset in Fig. \ref{figu2}. This reduction is found but then there is an unexpected over shoot at larger distances with a maximum of 19$\%$ at around $8\delta$.

\section{Strictly 2D metallic sheets as an illustrative example\label{metal}}

\begin{figure}
\includegraphics[width=7.0cm]{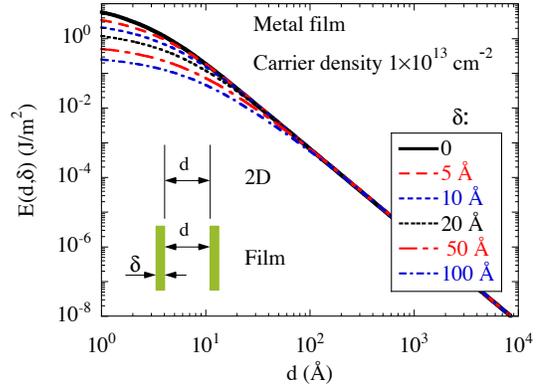}
\caption{(Color online) The attractive nonretarded interaction energy per unit area between two metal films as function of separation. They all have the projected 2D carrier density $1 \times {10^{13}}$cm$^{-2}$.}
\label{figu3}
\end{figure}

For 2D metallic sheets we use the zero-temperature RPA expression:
\begin{equation}
\begin{array}{l}
{\alpha ^{2D}}\left( {Q,iW} \right) = \\
 \quad \quad \quad \quad\frac{y}{Q}\left\{ {1 - \left[ {\sqrt {{{\left( {{Q^4} - {W^2} - {Q^2}} \right)}^2} + {{\left( {2W{Q^2}} \right)}^2}} } \right.} \right.\\
\quad \quad \quad \quad \quad \quad \quad \quad \quad  + \left. {{{\left. {{{\left( {{Q^4} - {W^2} - {Q^2}} \right)}^2}} \right]}^{1/2}}/{Q^2}} \right\},
\end{array}
\label{equ24}
\end{equation}
where
\begin{equation}
\begin{array}{l}
y = \frac{{m{e^2}}}{{{\hbar ^2}{k_F}}};\;W = \frac{{\hbar \omega }}{{4{E_F}}};\;Q = \frac{k}{{2{E_F}}};\\
{k_F} = \sqrt {2\pi {n^{2D}}} ;\;{E_F} = \frac{{{\hbar ^2}k_F^2}}{{2m}}.
\end{array}
\label{equ25}
\end{equation}
The result for the vdW Casimir interaction energy is shown in Fig. \ref{figu3} for both strictly 2D metal sheets and for a series of films with different film thickness. For large separations the interaction follows a fractional power law \cite{SerBjo,Bos,Dobson, Ser4}. At the small separation end of the figure the interaction weakens due to spatial dispersion. 

\begin{figure}
\includegraphics[width=7.0cm]{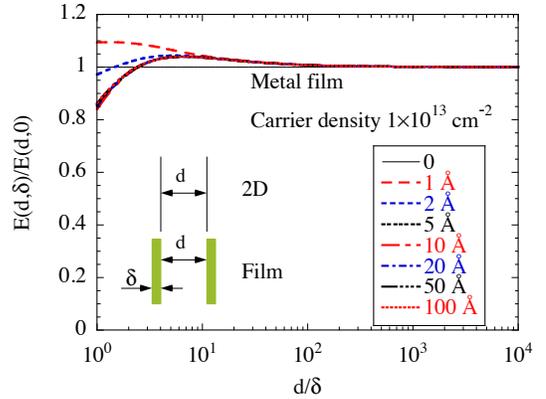}
\caption{(Color online) The ratio between the film result and the 2D sheet result for the interaction energy between two metal films as function of the ratio between the separation and the film thickness. All films have the projected 2D carrier density $1 \times {10^{13}}$cm$^{-2}$. Note that all curves with thickness of 5{\AA}  or larger fall more or less on the same curve.}
\label{figu4}
\end{figure}

In the metal film case we cannot produce a universal curve for the correction factor. Instead we give in Fig. \ref{figu4} the correction factor for a series of films with different film thickness. However for larger film thickness the results approach a universal curve. We see that the overall behavior is similar to that in the graphene case but the over shooting is here smaller.

\section{Non planar structures\label{nonplanar}}
In this section we discuss how one may proceed in non-planar structures, One may, e.g., have a cylinder or a sphere coated by a graphene or graphene-like film. The spatial dispersion complicates things. In these structures the momentum, $k$, is no longer a good quantum number for the normal modes. However, the problems  are often dominated by long wavelengths. The 3D polarizability for a graphene film in the long wavelength limit is
\begin{equation}
{\alpha ^{3D}}\left( {i\omega } \right) \approx \frac{{\pi {e^2}}}{{\hbar \delta \omega }},
\label{equ26}
\end{equation}
Fortunately the wave number is now absent from the expression and nothing hinders the use of this expression in non-planar structures.

For the metal film in the long wavelength limit
\begin{equation}
{\alpha ^{2D}}\left( {Q,iW} \right) \approx \frac{{yQ}}{{2{W^2}}},
\label{equ27}
\end{equation}
and hence
\begin{equation}
{\alpha ^{3D}}\left( {Q,iW} \right) \approx \frac{{yQ}}{{{W^2}q\delta }} = \frac{{4\pi {n^{2D}}{e^2}}}{{m{\omega ^2}\delta }} = \frac{{4\pi {n^{3D}}{e^2}}}{{m{\omega ^2}}} = \frac{{\omega _{pl}^2}}{{{\omega ^2}}},
\label{equ28}
\end{equation}
i.e., the ordinary Drude result for a 3D metal.
Once again the wave number is absent from the expression and nothing hinders the use of this expression in non-planar structures.

\section{Summary and conclusions\label{sum}}
We have shown a way to modify the dielectric function when a film of finite thickness is used to simulate a strictly 2D sheet or vice versa. We have used both a graphene sheet and a strictly 2D metal sheet as illustrations. We further pointed out how to proceed in the case of non-planar structures as to avoid problems caused by spatial dispersion.



%
%

%

\begin{acknowledgments}
We are grateful for financial support from the Swedish Research Council, VR Contract No. 70529001.
\end{acknowledgments}


\end{document}